# Distributed Model Predicted Control of Multi-agent Systems with Applications to Multi-vehicle Cooperation

Yougang Bian, *Member, IEEE*, Changkun Du, *Member, IEEE*, Manjiang Hu,

Haikuo Liu, *Member*

*Abstract*—This paper proposes a distributed model predicted control (DMPC) approach for consensus control of multi-agent systems (MASs) with linear agent dynamics and bounded control input constraints. Within the proposed DMPC framework, each agent exchanges assumed state trajectories with neighbors and solves a local open-loop optimization problem to obtain the optimal control input. In the optimization problem, a discrete-time consensus protocol is introduced into update law design for assumed terminal states, with which asymptotic consensus of assumed terminal states and recursive feasibility are rigorously proved. Together with the optimal cost function, an infinite series of cost-to-go functions is introduced into the design of a Lyapunov function, with which closed-loop asymptotic consensus is finally proved. Two applications including cooperation of autonomous underwater vehicles (AUVs) and connected and automated vehicles (CAVs) are used to validate the effectiveness of the proposed DMPC approach.

## I. Introduction

Distributed cooperative control of multi-agent systems (MASs) has emerged in the last few decades due to the popularization of mobile communication and computation techniques. Among the existing topics on related studies, the distributed consensus control technique has attracted lots of attention and has been applied to many industrial applications, e.g., platoon control of connected and automated vehicles (CAVs) [4][13], automated intersection management [2][5], and cooperation of autonomous underwater vehicles (AUVs) [6][7].

Among the existing studies on coordinated control of MASs, distributed model predictive control (DMPC) or distributed receding horizon control (DRHC) is an attractive approach since it can explicitly address dynamics nonlinearities and time-domain constraints while achieving specific performance optimization. For example, a DRHC protocol was proposed in [17] to achieve consensus of MASs with first-order dynamics for both the finite and infinite horizon cases. An inverse optimality approach was proposed in [16] for distributed receding horizon consensus of MASs with semi-stable and unstable general linear dynamics. When it comes to real applications, the DMPC technique has been used in [4] to achieve internal stable and string stable vehicle platoon control under the leader predecessor following (LPF) topology. Coupled cost functions and decoupled dynamics are considered in [4] to guarantee internal stability while coupled constraints were added to guarantee string stability. Besides traditional string stability, the concept of $\gamma$-gain stability was further introduced in [18] for vehicle platoons with the bidirectional (BD) topology. On the basis of [4], the LPF topology was further extended to the case of unidirectional communication topology in [8] and directed and acyclic graph (DAG) in [1] to account for more general communication topologies. By further considering of the unreliability of communication channels, the common Lyapunov technique (CLF) was used in [1] to extend [8] to the case of switching communication topology.

This paper studies DMPC for MASs with linear dynamics and bounded control input constraints. The contribution of this paper includes:

1) A discrete-time consensus protocol is introduced into the design of update law for assumed terminal states to guarantee asymptotic terminal consensus. Compared with [1][8], the proposed update law guarantees recursive feasibility rigorously during the consensus process of assumed terminal states;

2) An infinite series of cost-to-go functions is introduced into the Lyapunov function to cancel the non-zero assumed terminal errors for closed-loop asymptotic consensus analysis. Compared with [1][4][8], the proposed Lyapunov function relaxes the requirement on the communication topology to the

This study was supported by Key R&D Program of Hunan Province, China with 2019GK2161 and 2019GK2151, and State Key Laboratory of Advanced Design and Manufacturing for Vehicle Body with No. 61775006. (*Yougang Bian and Changkun Du contributed equally to this work. Corresponding author: Manjiang Hu.*)

Y. Bian and M. Hu are with the State Key Laboratory of Advanced Design and Manufacturing for Vehicle Body, College of Mechanical and Vehicle Engineering, Hunan University, Changsha, 410082, China (e-mail: byg10@foxmail.com, manjiang_h@hnu.edu.cn).

C. Du is with Beijing Advanced Innovation Center for Intelligent Robots and Systems, School of Mechatronical Engineering, Beijing Institute of Technology, Beijing 100081, China (e-mail: duchangkun88@gmail.com).

H. Liu is with the School of Mechatronical Engineering, Beijing Institute of Technology, Beijing 100081, China (e-mail: foreverlhk1220@126.com).

case of a spanning tree rooted at the leader.

The rest of this paper is organized as follows. Section II formulates the studied problem. Section III designs the DMPC controller. Section IV presents the analysis for the closed-loop system. Section V validates the proposed theorems with two application cases while Section VI concludes the paper.

*Notations:* The fields of integers, real numbers, and complex numbers are denoted by $\mathbb{N}$, $\mathbb{R}$, and $\mathbb{C}$ respectively. Define $\|x\|_P = \sqrt{x^T P x}$, where $x \in \mathbb{R}^n$ and $P \in \mathbb{R}^{n \times n}$ is a positive definite matrix. Denote by $\otimes$ the Kronecker product. Denote by $x(k|t)$ the value of $x(t+k)$ predicted at time $t$. For an arbitrary $N \in \mathbb{N}^+$, define $\mathbb{F} = \{1, 2, \ldots, N\}$. For arbitrary $N_1, N_2 \in \mathbb{N}$ satisfying $N_1 < N_2$, define $\mathbb{K}_{N_1:N_2} = \{N_1, N_1 + 1, \ldots, N_2 - 1, N_2\}$. For an arbitrary function $f(\cdot)$, abbreviate $f(x_1(k|t), \ldots, x_m(k|t))$ and $f(x_1(N_1|t), \ldots, x_1(N_2|t), \ldots, x_m(N_1|t), \ldots x_m(N_2|t))$ as $f(x_1, \ldots, x_m; k|t)$ and $f(x_1, \ldots, x_m; \mathbb{K}_{N_1:N_2}|t)$, respectively. For an arbitrary symmetric matrix $P \in \mathbb{R}^{n \times n}$, denote by $\lambda_i(P), i \in \{1, 2, \ldots, n\}$ its eigenvalues and $\lambda_M(P)$ its maximum eigenvalue. For an arbitrary set $\mathbb{S}$, denote by $|\mathbb{S}|$ its cardinality. A diagonal matrix with diagonal entries $d_1, d_2, \ldots, d_N \in \mathbb{R}$ is denoted by $\text{diag}\{d_1, d_2, \ldots, d_N\}$. Given a constant $c \in \mathbb{R}^+$, a continuous function $f: [0, c) \to [0, \infty)$ is called a class $\mathcal{K}$ function if it is strictly increasing and $f(0) = 0$.

## II. PROBLEM FORMULATION

This section formulates the consensus problem. We first model the agent dynamics and communication topology, and then present the control objective.

### A. Agent dynamics modeling

Consider a group of $N \in \mathbb{N}^+$ following agents with dynamics given as
$$x_i(t+1) = Ax_i(t) + Bu_i(t), \quad u_i(t) \in \mathbb{U}_i, \quad (1)$$
where $i \in \mathbb{F}$, $A \in \mathbb{R}^{n \times n}$, $B \in \mathbb{R}^{n \times m}$, $x_i(t) \in \mathbb{R}^n$ is the agent state at time $t$, $u_i(t) \in \mathbb{R}^m$ is the control input, and $\mathbb{U}_i = \{u_i | u_{i,m} \leq u_i \leq u_{i,M}\}$ is the admissible control input set with the origin as its interior point. For system (1), we make the following assumption.

*Assumption 1:* $(A, B)$ is controllable.

The dynamics of the leading agent (denoted by agent 0) is given as
$$x_0(t+1) = Ax_0(t), \quad (2)$$
where $x_0(t) \in \mathbb{R}^n$ is the leader state at time $t$.

*Remark 1:* The considered leader has no control inputs (or has a zero control input), as is also considered in [1][8]. Interested readers are referred to [15] for the case of a dynamic leader.

### B. Communication topology modeling

To model the communication among the following agents, define a directed graph $\mathcal{G} = \{\mathcal{V}, \mathcal{E}, \mathcal{A}\}$, where $\mathcal{V} = \{\mathcal{V}_1, \mathcal{V}_2, \ldots, \mathcal{V}_N\}$ is the set of nodes (following agents), $\mathcal{E} \subseteq \mathcal{V} \times \mathcal{V}$ is the set of edges (communication links), and $\mathcal{A} = [a_{ij}] \in \mathbb{R}^{N \times N}$ is the adjacency matrix. In $\mathcal{A}$, $a_{ij}$ equals 1 if agent $i$ can obtain the information of agent $j$, or 0 otherwise. Here we assume no self-loop, so $a_{ii} = 0$. Then we further define a degree matrix $\mathcal{D} = \text{diag}\{d_1, d_2, \ldots, d_N\}$, where $d_i = \sum_{j=1}^{N} a_{ij}$, and a Laplacian matrix $\mathcal{L} = \mathcal{D} - \mathcal{A}$. A directed graph $\mathcal{G}$ is said to contain a spanning tree if there exists a tree-type subgraph of $\mathcal{G}$ that includes all of its nodes.

To model the communication among the leader and the following agents, define a pinning matrix $\mathcal{B} = \text{diag}\{b_1, b_2, \ldots, b_N\}$, where $b_i$ equals 1 if agent $i$ can obtain the information of the leader, or 0 otherwise. Then the augmented graph containing the leader and the following agents is denoted by $\bar{\mathcal{G}}$. We further define $\mathcal{D}_\mathcal{B} = \mathcal{D} + \mathcal{B}$ and $\mathcal{L}_\mathcal{B} = \mathcal{L} + \mathcal{B}$.

Based on the above modeling, the in- and out-neighbor sets of agent $i$ are respectively defined as
$$\mathbb{N}_i = \{j \in \mathbb{F}, j \neq i | a_{ij} = 1\},$$
$$\mathbb{O}_i = \{j \in \mathbb{F}, j \neq i | a_{ji} = 1\},$$
the pinning set of agent $i$ is defined as
$$\mathbb{P}_i = \begin{cases} \{0\}, & \text{if } b_i = 1, \\ \emptyset, & \text{if } b_i = 0, \end{cases}$$
and the pinning in-neighbor set of agent $i$ is defined as
$$\mathbb{I}_i = \mathbb{N}_i \cup \mathbb{P}_i,$$
which describes where agent $i$ can obtain information.

As is done in [1][8], we make the following assumption on the communication topology, which yields Lemma 1.

*Assumption 2:* $\bar{\mathcal{G}}$ contains a directed spanning tree rooted at the leader.

*Lemma 1 [8]:* If Assumption 2 holds, $\mathcal{D}_\mathcal{B}$ is invertible and $|\lambda_i(\mathcal{D}_\mathcal{B}^{-1}\mathcal{A})| < 1, \forall i \in \mathbb{F}$.

This study assumes no communication time delays and packet drops. Interested readers are referred to [19][20] and [1] for the case of unreliable communication channels.

### C. Control objective formulation

The control objective of leader-following consensus is given below:
$$\lim_{t \to +\infty} x_i(t) = x_0(t), \quad \forall i \in \mathbb{F}, \quad (3)$$
which requires the following agents to track the leader asymptotically.

## III. DMPC CONTROLLER DESIGN

This section designs the DMPC controller. We first design an open-loop optimization problem, and then present the terminal state update law, after which the DMPC algorithm is designed.

### A. Open-loop optimization problem design

Denote by $N_p$ the predictive horizon. Then we design the following three types of trajectories in the predictive horizon:

1) $x_i^p(k|t)$, $k \in \mathbb{K}_{0:N_p}$ and $u_i^p(k|t)$, $k \in \mathbb{K}_{0:N_p-1}$: predicted state and control input trajectories;

2) $x_i^a(k|t)$, $k \in \mathbb{K}_{0:N_p}$ and $u_i^a(k|t)$, $k \in \mathbb{K}_{0:N_p-1}$: assumed state and control input trajectories;

3) $x_i^*(k|t)$, $k \in \mathbb{K}_{0:N_p}$ and $u_i^*(k|t)$, $k \in \mathbb{K}_{0:N_p-1}$: optimal state and control input trajectories.

Then the open-loop optimization problem is designed as follows:

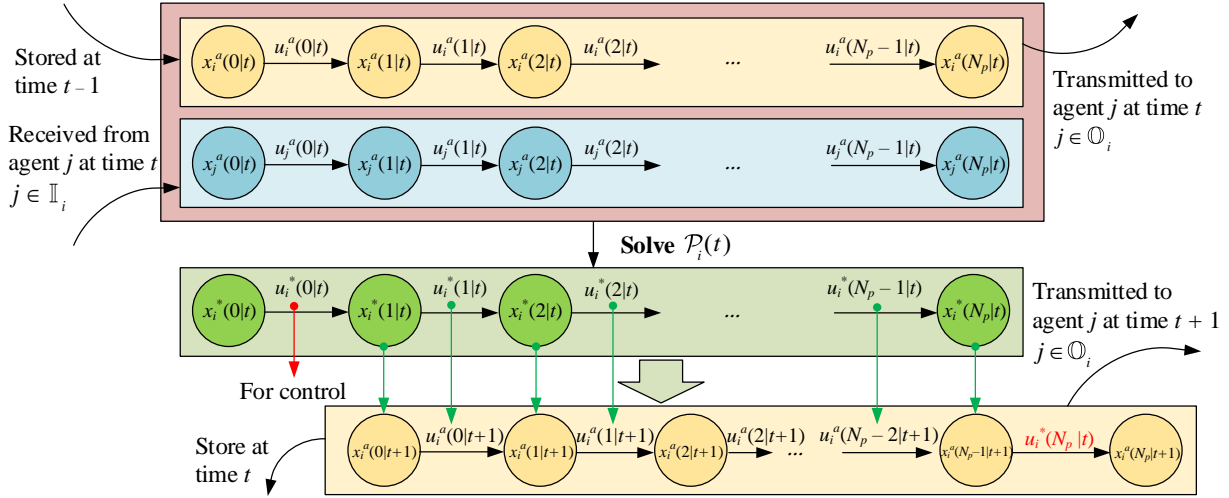

Fig. 1 Illustration of the DMPC algorithm.

*Problem $\mathcal{P}_i(t)$:*

$$\min_{u_i^p(k|t), k\in\mathbb{K}_{0:N_p-1}} J_i\left(x_i^p, u_i^p, x_i^a, x_{j\in\mathbb{I}_i}^a; \mathbb{K}_{0:N_p-1}|t\right) \tag{4}$$
$$= \sum_{k=0}^{N_p-1} l_i\left(x_i^p, u_i^p, x_i^a, x_{j\in\mathbb{I}_i}^a; k|t\right)$$

*subject to:*

$$x_i^p(0|t) = x_i(t), \tag{5}$$
$$x_i^p(k+1|t) = Ax_i^p(k|t) + Bu_i^p(k|t), \ k \in \mathbb{K}_{0:N_p-1}, \tag{6}$$
$$u_i^p(k|t) \in \mathbb{U}_i, \ k \in \mathbb{K}_{0:N_p-1}, \tag{7}$$
$$x_i^p(N_p|t) = x_i^a(N_p|t), \tag{8}$$

where

$$\begin{aligned}&l_i\left(x_i^p, u_i^p, x_i^a, x_{j\in\mathbb{I}_i}^a; k|t\right) \\ &= \|u_i^p(k|t)\|_{R_i} + \|x_i^p(k|t) - x_i^a(k|t)\|_{F_i} \\ &+ \sum_{j\in\mathbb{I}_i} \|x_i^p(k|t) - x_j^a(k|t)\|_{G_i}.\end{aligned} \tag{9}$$

In problem $\mathcal{P}_i(t)$, cost function (4) and (9) penalizes three terms: 1) the control input with weight matrix $R_i \geq 0$, 2) the deviation between the predicted and assumed states of agent $i$ (or *self-deviation* for short) with weight matrix $F_i \geq 0$, and 3) the deviation between agent $i$'s predicted states and agent $j$'s assumed states (or *neighbor-deviation* for short) with weight matrix $G_i \geq 0$.

In problem $\mathcal{P}_i(t)$, constraints (5), (6), and (7) represent the initial state constraint, agent dynamics constraint, and control input constraint, respectively, while constraint (8) is a terminal state constraint. Here the assumed terminal state $x_i^a(N_p|t)$ in (8) is updated in each control loop, and the detailed update law is designed in the following subsection.

*B. Update law design for assumed terminal states*

The update law for assumed terminal state $x_i^a(N_p|t)$ is designed here to guarantee both recursive feasibility and terminal consensus. From (8) we have $x_i^*(N_p|t) = x_i^a(N_p|t)$. Then, inspired by [12], we introduce the following discrete-time consensus protocol to the update law design for agent $i$:

$$x_i^a(N_p|t+1) = Ax_i^a(N_p|t) + Bu_i^*(N_p|t), \tag{10}$$

where

$$u_i^*(N_p|t) = \frac{1}{|\mathbb{I}_i|} K \sum_{j\in\mathbb{I}_i} \left(x_j^a(N_p|t) - x_i^a(N_p|t)\right). \tag{11}$$

Note that $u_i^*(N_p|t)$ in (11) actually does not exist at time $t$ but is defined here for notation simplicity. Moreover, $K$ in (11) is given as

$$K = -(B^T PB + I)^{-1} B^T PA, \tag{12}$$

where $P > 0$ is the solution of the following modified algebraic Riccati equation (MARE):

$$P = -(1-\delta^2)A^T PB(B^T PB + I)^{-1} B^T PA \\ + A^T PA + Q, \tag{13}$$

where $Q > 0$, $\delta \in \mathbb{R}$ are constant parameters. Then we have the following lemma on the existence of $P$, which will be used in terminal consensus analysis (see Section IV.A).

*Lemma 2 [12]:* Suppose that Assumption 1 holds. If $A$ has no eigenvalues with magnitude larger than 1, then for any $0 < \delta < 1$, MARE (13) has a unique solution $P > 0$, and $A - (1-\sigma)BK$ is Schur stable for any $\sigma \in \mathbb{C}$ satisfying $|\sigma| < \delta$. Moreover, if $A$ has at least one eigenvalue with magnitude larger than 1 and $B$ is of rank one, then for any $0 < \delta < \frac{1}{\prod_i |\lambda_i^u(A)|}$, where $\lambda_i^u(A)$ denotes the unstable eigenvalue of $A$, MARE (13) has a unique solution $P > 0$, and $A - (1-\sigma)BK$ is Schur stable for any $\sigma \in \mathbb{C}$ satisfying $|\sigma| < \delta$.

*Remark 2:* Different from [1][8] where a terminal constraint is designed as $x_i^p(N_p|t) = \frac{1}{|\mathbb{I}_i|}\sum_{j\in\mathbb{I}_i} x_j^a(N_p|t)$ while the assumed terminal state is updated by $x_j^a(N_p|t+1) = Ax_j^*(N_p|t)$, the proposed terminal constraint (8) and update law for assumed terminal states in (10)-(13) take agent dynamics into account, which helps guarantee recursive feasibility as well as terminal consensus (see Sections IV.A and IV.B).

*C. DMPC algorithm design*

With the proposed open-loop optimization problem and update law for assumed terminal states, the DMPC algorithm is given in TABLE I and is further illustrated in Fig. 1. As shown in Fig. 1, agent $i$ receives its in-neighbors' assumed trajectories at time $t$, and then use them, together with its own assumed trajectory, to solve $\mathcal{P}_i(t)$. The first control input of the optimal solution is used for the control of agent $i$ at time $t$, while the others are combined with the updated terminal control input to generate the assumed trajectory of time $t + 1$.

## IV. CLOSED-LOOP SYSTEM ANALYSIS

This section presents the analysis for the closed-loop system. We first analyze the consensus of assumed terminal states, and then prove the recursive feasibility. The asymptotic consensus of the closed-loop system is proved finally.

### A. Terminal consensus analysis

Terminal consensus requires that the terminal states of the followers' assumed trajectories achieve consensus with that of the leader. Fig. 2 shows an example for terminal consensus, where the followers' assumed trajectories intersect in one point with each other at the terminal in the 2 dimensional phase plane. Terminal consensus facilitates the analysis of recursive feasibility and asymptotic stability in the following subsections, and is also considered in [1][8].

TABLE I
DMPC ALGORITHM

**Initialization**: for all agents $i \in \mathbb{F}$,
1. Prepare $u_i^a(k|0), k \in \mathbb{K}_{0:N_p-1}$ and $x_i^a(k|0), k \in \mathbb{K}_{0:N_p}$:
$$u_i^a(k|0) \in \mathbb{U}_i, \quad k \in \mathbb{K}_{0:N_p-1}$$
$$x_i^a(k|0) = \begin{cases} x_i(0), & k = 0, \\ Ax_i^a(k-1|0) + Bu_i^a(k-1|0), & k \in \mathbb{K}_{1:N_p}. \end{cases}$$

**At time** $t \geq 0$: for all agents $i \in \mathbb{F}$,
1. Transmit $x_i^a(k|t), k \in \mathbb{K}_{0:N_p}$ to agents $j \in \mathbb{O}_i$ through communication;
2. Receive $x_j^a(k|t), k \in \mathbb{K}_{0:N_p}$ from agents $j \in \mathbb{I}_i$ through communication;
3. Solve $\mathcal{P}_i(t)$ and obtain $u_i^*(k|t), k \in \mathbb{K}_{0:N_p-1}$ and $x_i^*(k|t), k \in \mathbb{K}_{0:N_p}$;
4. Use $u_i^*(0|t)$ for agent $i$'s control;
5. Prepare $u_i^a(k|t+1), k \in \mathbb{K}_{0:N_p-1}$ and $x_i^a(k|t+1), k \in \mathbb{K}_{0:N_p}$:
$$u_i^a(k|t+1) = \begin{cases} u_i^*(k+1|t), & k \in \mathbb{K}_{0:N_p-2}, \\ u_i^*(N_p|t) = \frac{1}{|\mathbb{I}_i|} K \sum_{j \in \mathbb{I}_i} \left(x_j^a(N_p|t) - x_i^a(N_p|t)\right), & k = N_p - 1, \end{cases}$$
$$x_i^a(k|t+1) = \begin{cases} x_i^*(k+1|t), & k \in \mathbb{K}_{0:N_p-1}, \\ Ax_i^a(N_p-1|t+1) + Bu_i^a(N_p-1|t+1), & k = N_p. \end{cases}$$

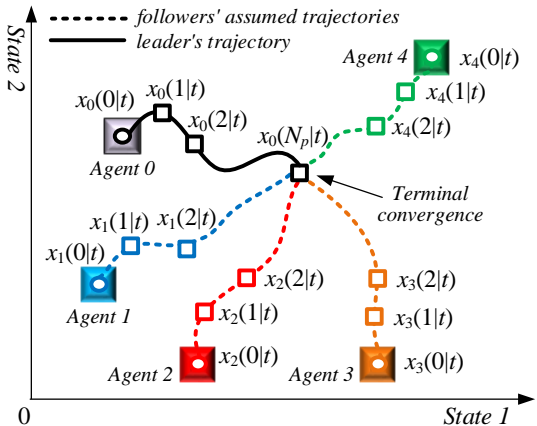

Fig. 2  Illustration of terminal consensus.

Define the assumed terminal error as
$$e_i^a(N_p|t) = x_i^a(N_p|t) - x_0(N_p|t).$$
Then according to (2) and (10), we have
$$e_i^a(N_p|t) = Ae_i^a(N_p|t-1) + Bu_i^a(N_p|t-1). \quad (14)$$
Define stacked errors as
$$E^a(N_p|t) = \left[e_1^a(N_p|t)^T, e_2^a(N_p|t)^T, \ldots, e_N^a(N_p|t)^T\right]^T$$
and stacked control inputs as
$$U^a(N_p|t) = \left[u_1^a(N_p|t)^T, u_2^a(N_p|t)^T, \ldots, u_N^a(N_p|t)^T\right]^T,$$
then according to (11), we have
$$U^a(N_p|t-1) = -(\mathcal{D}_\mathcal{B}^{-1}\mathcal{L}_\mathcal{B} \otimes K) \cdot E^a(N_p|t-1). \quad (15)$$
Then combining (14) and (15) yields
$$E^a(N_p|t) = (I_N \otimes A - \mathcal{D}_\mathcal{B}^{-1}\mathcal{L}_\mathcal{B} \otimes BK) \cdot E^a(N_p|t-1). \quad (16)$$

Now we present the following theorem and prove that $E^a(N_p|t)$ will converge to zero asymptotically.

*Theorem 1:* For a multi-agent system with dynamics (1)-(2) and DMPC controller (4)-(13), suppose Assumption 1 and Assumption 2 hold. For $\delta$ in MARE (13), if it holds that
$$\max_{i \in \mathbb{F}} |\lambda_i(\mathcal{D}_\mathcal{B}^{-1}\mathcal{A})| < \delta < 1, \max_j |\lambda_j(A)| \leq 1, \quad (17)$$
or
$$\max_{i \in \mathbb{F}} |\lambda_i(\mathcal{D}_\mathcal{B}^{-1}\mathcal{A})| < \delta < \frac{1}{\prod_i |\lambda_i^u(A)|}, \quad (18)$$
$$\max_j |\lambda_j(A)| > 1,$$
then the assumed terminal errors (16) converge to zero asymptotically.

*Proof:* Since $\mathcal{D}_\mathcal{B}^{-1}\mathcal{L}_\mathcal{B} = I_N - \mathcal{D}_\mathcal{B}^{-1}\mathcal{A}$, the eigenvalues of $\mathcal{D}_\mathcal{B}^{-1}\mathcal{L}_\mathcal{B}$ become $1 - \lambda_i(\mathcal{D}_\mathcal{B}^{-1}\mathcal{A}), i \in \mathbb{F}$. Then we know there exists an invertible matrix $W$ such that
$$W^{-1}\mathcal{D}_\mathcal{B}^{-1}\mathcal{L}_\mathcal{B} W$$
$$= \Lambda \coloneqq \begin{bmatrix} 1 - \lambda_1(\mathcal{D}_\mathcal{B}^{-1}\mathcal{A}) & \cdots & * \\ \vdots & \ddots & \vdots \\ 0 & \cdots & 1 - \lambda_N(\mathcal{D}_\mathcal{B}^{-1}\mathcal{A}) \end{bmatrix}.$$
Introduce an invertible state transformation
$$\tilde{E}^a(N_p|t) = (W^{-1} \otimes I_n)E^a(N_p|t),$$
then (16) can be transformed into
$$\tilde{E}^a(N_p|t) = \tilde{A} \cdot \tilde{E}^a(N_p|t-1),$$
where
$$\tilde{A} = I_N \otimes A - \Lambda \otimes BK$$
$$= \begin{bmatrix} A - (1 - \lambda_1(\mathcal{D}_\mathcal{B}^{-1}\mathcal{A}))BK & \cdots & * \\ \vdots & \ddots & \vdots \\ 0 & \cdots & A - (1 - \lambda_N(\mathcal{D}_\mathcal{B}^{-1}\mathcal{A}))BK \end{bmatrix}.$$
When (17) or (18) holds, since Assumption 1 holds, then according to Lemma 2, $A - (1 - \lambda_i(\mathcal{D}_\mathcal{B}^{-1}\mathcal{A}))BK$ is Schur stable, which indicates that $\tilde{A}$ is also Schur stable. This completes the proof. ∎

*Remark 3:* In [1][8], the communication topology is required to be unidirectional or to be a directed acyclic graph, which guarantees that the assumed terminal states achieve consensus in finite time. Different from [1][8], this study relaxes the requirement on communication topology to the case of containing a spanning tree rooted at the leader. This brings great challenge to the proof of closed-loop consensus (see Section C).

*Remark 4:* In the case of $|\lambda_i(A)| < 1, \forall i \in \mathbb{F}$, since Assumption 2 holds, according Lemma 1, we have $|\lambda_i(\mathcal{D}_B^{-1}\mathcal{A})| < 1, \forall i \in \mathbb{F}$. Then there always exists a constant $\delta$ such that $\max_{i \in \mathbb{F}}|\lambda_i(\mathcal{D}_B^{-1}\mathcal{A})| < \delta < 1$, i.e., (17) holds. In the case of $|\lambda_i(A)| < 1, \exists i \in \mathbb{F}$, it is required that $\max_{i \in \mathbb{F}}|\lambda_i(\mathcal{D}_B^{-1}\mathcal{A})| < \frac{1}{\prod_i |\lambda_i^u(A)|}$ such that $\delta$ exists.

### B. Recursive feasibility analysis

Recursive feasibility requires the open-loop optimization problem $\mathcal{P}_i(t)$ to be always feasible provided initial feasibility. Here we directly present the following theorem on recursive feasibility.

*Theorem 2:* For a multi-agent system with dynamics (1)-(2) and DMPC controller (4)-(13), suppose Assumption 1 and Assumption 2 hold, and (17) or (18) holds. Given the initial feasibility of $\mathcal{P}_i(t_0), \forall i \in \mathbb{F}$, there exists a non-empty set $\mathbb{E}$ such that when $E^a(N_p|t_0) \in \mathbb{E}$, $\mathcal{P}_i(t_0 + 1), \forall i \in \mathbb{F}$ is recursive feasible.

*Proof:* According to the DMPC algorithm, we only need to prove that there always exists a set of $u_i^p(k|t+1), k \in \mathbb{K}_{0:N_p-1}$ satisfying constraint (7) given the feasibility of $\mathcal{P}_i(t), \forall i \in \mathbb{F}$. Substituting (16) into (15) yields
$$U^a(N_p|t) = -(\mathcal{D}_B^{-1}\mathcal{L}_B \otimes K)\bar{A}^t \cdot E^a(N_p|0),$$
where $\bar{A} = I_N \otimes A - \mathcal{D}_B^{-1}\mathcal{L}_B \otimes BK$. According to Theorem 1, since $W$ is invertible, we know $\bar{A}$ is similar to $\tilde{A}$, so $\bar{A}$ is also Schur stable. Then we know $U^a(N_p|t)$ is asymptotically convergent, i.e., $\lim_{t \to +\infty} U^a(N_p|t) = 0$. Then there exists a neighborhood of the origin, denoted by $\mathbb{E}$, such that when $E^a(N_p|0) \in \mathbb{E}$, it holds that all the elements of $U^a(N_p|t)$ satisfy constraint (7) since the origin is an interior point of $\mathbb{U}_i, i \in \mathbb{F}$. Therefore, we know that $u_i^*(k|t), k \in \mathbb{K}_{1:N_p}$ is a feasible solution to $\mathcal{P}_i(t+1)$. This completes the proof. ∎

*Remark 5:* Different from [1][8], which guarantee recursive feasibility only when the assumed terminal errors converge to zero, Theorem 2 only requires the initial assumed terminal errors to lie in a neighborhood $\mathbb{E}$ of the origin. This expands the feasibility of the proposed DMPC controller. In practice, the accurate formulation of $\mathbb{E}$ is difficult, and it can be estimated through trial-and-error by continuously contracting the norm of $E^a(N_p|0)$ and checking the satisfaction of the corresponding control input constraint.

### C. Closed-loop consensus analysis

Closed-loop consensus requires the closed-loop consensus errors to converge to zero asymptotically. Before presenting the analysis, the following lemmas are given first.

*Lemma 3 [11]:* Consider a system with dynamics
$$x(t+1) = f(x(t)), t \geq 0, x(0) = x_0,$$
where $f$ is continuous with respect to $x$ and $f(0) = 0$. the system is asymptotically stable if there exists a positive bounded function $V(x)$ satisfying:
(a) $V(0) = 0$, and $V(x) > 0$ for any $x \neq 0$;
(b) $V(x)$ is continuous in the neighbor zone of the equilibrium $x = 0$;
(c) along the system trajectory it holds
$$V(x(t+1)) - V(x(t)) \leq -\alpha_0(\|x(t)\|),$$
where $\alpha_0(\cdot)$ is a class $\mathcal{K}$ function.

*Lemma 4 [10]:* For arbitrary $x_1, x_2 \in \mathbb{R}^n$ and positive definite $P \in \mathbb{R}^{n \times n}$, it holds that
$$\|x_1\|_P + \|x_2\|_P \geq \|x_1 + x_2\|_P.$$

*Lemma 5 [9]:* For arbitrary $x_1, x_2, \ldots, x_m \in \mathbb{R}$, it holds that
$$\left(\sum_{i=1}^m |x_i|\right)^2 \leq m \sum_{i=1}^m x_i^2.$$

Now we are ready to present the theorem on asymptotic consensus.

*Theorem 3:* The closed-loop system achieves asymptotic consensus objective (3) if
$$F_i \geq |\mathbb{O}_i| \cdot \sum_{j \in \mathbb{O}_i} G_j. \tag{19}$$

*Proof:* Consider the following Lyapunov candidate
$$V(t) = \sum_{i=1}^N V_i(t) = \sum_{i=1}^N (J_i^*(t) + q_i^*(t)),$$
where $J_i^*(t)$ is the optimal cost function, i.e.,
$$J_i^*(t) = J_i\left(x_i^*, u_i^*, x_i^a, x_{j \in \mathbb{I}_i}^a; \mathbb{K}_{0:N_p-1}|t\right)$$
$$= \sum_{k=0}^{N_p-1} l_i(x_i^*, u_i^*, x_i^a, x_{j \in \mathbb{I}_i}^a; k|t),$$
and $q_i^*(t)$ is an infinity series of cost-to-go functions given as follows
$$q_i^*(t) = \sum_{l=t}^{+\infty} l_i(x_i^*, u_i^*, x_i^*, x_{j \in \mathbb{I}_i}^*; N_p|l).$$

It is clear that $V(t) \geq 0$, and the equality holds if and only if $e_i^*(k|t) = 0 \ \forall i \in \mathbb{F}$, which indicates that the closed-loop system achieves consensus. In addition, $V(t)$ is continuous around $e_i^*(k|t) = 0$.

According to Lemma 3, we first prove that $V(t)$ is bounded. It is obvious that $\sum_{i=1}^N J_i^*(t)$ given bounded initial states. Then we only need to consider the boundedness of $\sum_{i=1}^N q_i^*(t)$. We have

$$\sum_{i=1}^N q_i^*(t)$$
$$= \sum_{i=1}^N \sum_{l=t}^{+\infty} \left( \left\| \frac{1}{|\mathbb{I}_i|} K \sum_{j \in \mathbb{I}_i} \left(e_i^*(N_p|l) - e_j^*(N_p|l)\right) \right\|_{R_i} \right.$$
$$+ \sum_{j \in \mathbb{I}_i} \left\| e_i^*(N_p|l) - e_j^*(N_p|l) \right\|_{G_i} \right) \tag{20}$$

$$= \sum_{i=1}^{N} \sum_{l=t}^{+\infty} \left( \left\| \sum_{j \in \mathbb{I}_i} \left( e_i^*(N_p|l) - e_j^*(N_p|l) \right) \right\|_{\frac{K^T R_i K}{|\mathbb{I}_i|^2}} + \sum_{j \in \mathbb{I}_i} \left\| e_i^*(N_p|l) - e_j^*(N_p|l) \right\|_{G_i} \right).$$

Define
$$Q_{ij} = \left[ 0_n, 0_n, \ldots, \underbrace{I_n}_{ith}, \ldots, \underbrace{-I_n}_{jth}, \ldots, 0_n \right] \in \mathbb{R}^{n \times nN}.$$

Then
$$e_i^*(N_p|l) - e_j^*(N_p|l) = Q_{ij} E^*(N_p|l).$$

Further we have
$$\|e_i^*(N_p|l) - e_j^*(N_p|l)\|_{G_i}^2 = \|Q_{ij} \cdot E^*(N_p|l)\|_{G_i}^2$$
$$= E^*(N_p|l)^T Q_{ij}^T G_i Q_{ij} E^*(N_p|l) \qquad (21)$$
$$= E^*(N_p|0)^T (\bar{A}^l)^T Q_{ij}^T G_i Q_{ij} \bar{A}^l E^*(N_p|0),$$

and
$$\left\| \sum_{j \in \mathbb{I}_i} \left( e_i^*(N_p|l) - e_j^*(N_p|l) \right) \right\|_{\frac{K^T R_i K}{|\mathbb{I}_i|^2}}^2$$
$$= \left\| \left( \sum_{j \in \mathbb{I}_i} Q_{ij} \right) \cdot E^*(N_p|l) \right\|_{\frac{K^T R_i K}{|\mathbb{I}_i|^2}}^2 \qquad (22)$$
$$= \|\bar{A}^l E^*(N_p|0)\|_{\left(\sum_{j \in \mathbb{I}_i} Q_{ij}^T\right) \frac{K^T R_i K}{|\mathbb{I}_i|^2} \left(\sum_{j \in \mathbb{I}_i} Q_{ij}\right)}^2$$
$$= \|E^*(N_p|0)\|_{(\bar{A}^l)^T \left(\sum_{j \in \mathbb{I}_i} Q_{ij}^T\right) \frac{K^T R_i K}{|\mathbb{I}_i|^2} \left(\sum_{j \in \mathbb{I}_i} Q_{ij}\right) \bar{A}^l}^2.$$

Denote by
$$\lambda_{1i} = \lambda_M \left( \left( \sum_{j \in \mathbb{I}_i} Q_{ij}^T \right) \frac{K^T R_i K}{|\mathbb{I}_i|^2} \left( \sum_{j \in \mathbb{I}_i} Q_{ij} \right) \right)$$

and
$$\lambda_{2ij} = \lambda_M (Q_{ij}^T G_i Q_{ij})$$

the maximum eigenvalues of the two matrices. Then substituting (21)-(22) into (20) yields (23).

Since $|\lambda_M(\bar{A})| < 1$, we have $0 \leq \sqrt{\lambda_M(\bar{A}^T \bar{A})} < 1$, then the power series is convergent. This indicates the boundedness of $\sum_{i=1}^{N} q_i^*(t)$ as well as $V(t)$.

Next, we prove that $V(t)$ is monotonically decreasing. According to the DMPC algorithm, we have $x_i^a(k|t+1) = x_i^*(k+1|t)$, $x_{j \in \mathbb{I}_i}^a(k|t+1) = x_{j \in \mathbb{I}_i}^*(k+1|t), k \in \mathbb{K}_{0:N_p-1}$. Moreover, according to the proof of Theorem 2, $u_i^*(k|t), k \in \mathbb{K}_{1:N_p}$ is a feasible solution to $\mathcal{P}_i(t+1)$, and $x_i^*(k|t), k \in$ $\mathbb{K}_{1:N_p}$ is the corresponding stage state. Then we have

$$J_i^*(t+1) - J_i^*(t)$$
$$= \sum_{k=0}^{N_p-1} l_i(x_i^*, u_i^*, x_i^a, x_{j \in \mathbb{I}_i}^a; k|t+1)$$
$$- \sum_{k=0}^{N_p-1} l_i(x_i^*, u_i^*, x_i^a, x_{j \in \mathbb{I}_i}^a; k|t)$$
$$= \sum_{k=0}^{N_p-1} l_i(x_i^*, u_i^*, x_i^a, x_{j \in \mathbb{I}_i}^a; k|t+1)$$
$$- \sum_{k=0}^{N_p-1} l_i(x_i^*, u_i^*, x_i^a, x_{j \in \mathbb{I}_i}^a; k|t)$$
$$\leq \sum_{k=1}^{N_p} l_i(x_i^*, u_i^*, x_i^*, x_{j \in \mathbb{I}_i}^*; k|t) - \sum_{k=0}^{N_p-1} l_i(x_i^*, u_i^*, x_i^a, x_{j \in \mathbb{I}_i}^a; k|t),$$

where the optimality of $J_i^*(t+1)$ is used in the inequality. Further, we have

$$J_i^*(t+1) - J_i^*(t)$$
$$\leq l_i(x_i^*, u_i^*, x_i^*, x_{j \in \mathbb{I}_i}^*; N_p|t) - l_i(x_i^*, u_i^*, x_i^a, x_{j \in \mathbb{I}_i}^a; 0|t)$$
$$+ \sum_{k=1}^{N_p-1} \left( \|u_i^*(k|t)\|_{R_i} + \|x_i^*(k|t) - x_i^*(k|t)\|_{F_i} \right.$$
$$\left. + \sum_{j \in \mathbb{I}_i} \|x_i^*(k|t) - x_j^*(k|t) + \tilde{d}_{i,j}\|_{G_i} \right)$$
$$- \sum_{k=1}^{N_p-1} \left( \|u_i^*(k|t)\|_{R_i} + \|x_i^*(k|t) - x_i^a(k|t)\|_{F_i} \right.$$
$$\left. + \sum_{j \in \mathbb{I}_i} \|x_i^*(k|t) - x_j^a(k|t) + \tilde{d}_{i,j}\|_{G_i} \right)$$
$$= l_i(x_i^*, u_i^*, x_i^*, x_{j \in \mathbb{I}_i}^*; N_p|t) - l_i(x_i^*, u_i^*, x_i^a, x_{j \in \mathbb{I}_i}^a; 0|t)$$
$$+ \sum_{k=1}^{N_p-1} \left( -\|x_i^*(k|t) - x_i^a(k|t)\|_{F_i} \right.$$
$$+ \sum_{j \in \mathbb{I}_i} \|x_i^*(k|t) - x_j^*(k|t) + \tilde{d}_{i,j}\|_{G_i}$$
$$\left. - \sum_{j \in \mathbb{I}_i} \|x_i^*(k|t) - x_j^a(k|t) + \tilde{d}_{i,j}\|_{G_i} \right)$$
$$\leq l_i(x_i^*, u_i^*, x_i^*, x_{j \in \mathbb{I}_i}^*; N_p|t) - l_i(x_i^*, u_i^*, x_i^a, x_{j \in \mathbb{I}_i}^a; 0|t)$$
$$+ \sum_{k=1}^{N_p-1} \left( -\|x_i^*(k|t) - x_i^a(k|t)\|_{F_i} \right.$$
$$\left. + \sum_{j \in \mathbb{I}_i} \|x_j^*(k|t) - x_j^a(k|t)\|_{G_i} \right)$$

$$\begin{aligned}
&\sum_{i=1}^{N} q_i^*(t) \\
&= \sum_{i=1}^{N} \sum_{l=t}^{+\infty} \left( \sqrt{E^*(N_p|0)^T (\bar{A}^l)^T \left(\sum_{j \in \mathbb{I}_i} Q_{ij}^T\right) \frac{K^T R_i K}{|\mathbb{I}_i|^2} \left(\sum_{j \in \mathbb{I}_i} Q_{ij}\right) \bar{A}^l E^*(N_p|0)} + \sum_{j \in \mathbb{I}_i} \sqrt{E^*(N_p|0)^T (\bar{A}^l)^T Q_{ij}^T G_i Q_{ij} \bar{A}^l E^*(N_p|0)} \right) \\
&\leq \sum_{i=1}^{N} \sum_{l=t}^{+\infty} \left( \sqrt{\lambda_{1i} \cdot E^*(N_p|0)^T (\bar{A}^l)^T \bar{A}^l E^*(N_p|0)} + \sum_{j \in \mathbb{I}_i} \sqrt{\lambda_{2ij} \cdot E^*(N_p|0)^T (\bar{A}^l)^T \bar{A}^l E^*(N_p|0)} \right) \\
&\leq \sum_{i=1}^{N} \sum_{l=t}^{+\infty} \left( \sqrt{\lambda_{1i} \cdot \lambda_M^l(\bar{A}^T \bar{A}) \cdot E^*(N_p|0)^T E^*(N_p|0)} + \sum_{j \in \mathbb{I}_i} \sqrt{\lambda_{2ij} \cdot \lambda_M^l(\bar{A}^T \bar{A}) \cdot E^*(N_p|0)^T E^*(N_p|0)} \right) \qquad (23) \\
&= \sum_{l=t}^{+\infty} \left(\sqrt{\lambda_M(\bar{A}^T \bar{A})}\right)^l \cdot \sqrt{E^*(N_p|0)^T E^*(N_p|0)} \cdot \sum_{i=1}^{N} \left( \sqrt{\lambda_{1i}} + \sum_{j \in \mathbb{I}_i} \sqrt{\lambda_{2ij}} \right).
\end{aligned}$$

---

$$= l_i(x_i^*, u_i^*, x_i^*, x_{j \in \mathbb{I}_i}^*; N_p|t) - l_i(x_i^*, u_i^*, x_i^a, x_{j \in \mathbb{I}_i}^a; 0|t)$$
$$+ \sum_{k=1}^{N_p-1} \left( -\|x_i^*(k|t) - x_i^a(k|t)\|_{F_i} + \sum_{j \in \mathbb{O}_i} \|x_i^*(k|t) - x_i^a(k|t)\|_{G_j} \right),$$

where Lemma 4 is used in the second inequality. Then we have
$$V(t+1) - V(t)$$
$$= \sum_{i=1}^{N} \left(J_i^*(t+1) - J_i^*(t) + q_i^*(t+1) - q_i^*(t)\right)$$
$$\leq \sum_{i=1}^{N} \Bigg[ l_i(x_i^*, u_i^*, x_i^*, x_{j \in \mathbb{I}_i}^*; N_p|t) - l_i(x_i^*, u_i^*, x_i^a, x_{j \in \mathbb{I}_i}^a; 0|t)$$
$$+ \sum_{k=1}^{N_p-1} \left( -\|x_i^*(k|t) - x_i^a(k|t)\|_{F_i} + \sum_{j \in \mathbb{O}_i} \|x_i^*(k|t) - x_i^a(k|t)\|_{G_j} \right)$$
$$+ \sum_{l=t+1}^{+\infty} l_i(x_i^*, u_i^*, x_i^*, x_{j \in \mathbb{I}_i}^*; N_p|l) - \sum_{l=t}^{+\infty} l_i(x_i^*, u_i^*, x_i^*, x_{j \in \mathbb{I}_i}^*; N_p|l) \Bigg]$$
$$= \sum_{i=1}^{N} \Bigg[ -l_i(x_i^*, u_i^*, x_i^a, x_{j \in \mathbb{I}_i}^a; 0|t)$$
$$+ \sum_{k=1}^{N_p-1} \left( -\|x_i^*(k|t) - x_i^a(k|t)\|_{F_i} + \sum_{j \in \mathbb{O}_i} \|x_i^*(k|t) - x_i^a(k|t)\|_{G_j} \right) \Bigg].$$

When (19) holds, we have
$$\|x_i^*(k|t) - x_i^a(k|t)\|_{F_i}^2 \geq |\mathbb{O}_i| \cdot \sum_{j \in \mathbb{O}_i} \|x_i^*(k|t) - x_i^a(k|t)\|_{G_j}^2$$
$$\geq \left(\sum_{j \in \mathbb{O}_i} \|x_i^*(k|t) - x_i^a(k|t)\|_{G_j}\right)^2,$$

where Lemma 5 is used in the second inequality. Then we have
$$-\|x_i^*(k|t) - x_i^a(k|t)\|_{F_i} + \sum_{j \in \mathbb{O}_i} \|x_i^*(k|t) - x_i^a(k|t)\|_{G_j} \leq 0,$$

which further yields
$$V(t+1) - V(t) \leq -\sum_{i=1}^{N} l_i(x_i^*, u_i^*, x_i^a, x_{j \in \mathbb{I}_i}^a; 0|t)$$
$$\leq -\sum_{i=1}^{N} \sum_{j \in \mathbb{I}_i} \|x_i(t) - x_j(t)\|_{G_i} \leq 0,$$

where the equality holds if and only if
$$x_i^*(k|t) = x_i^a(k|t), \forall i \in \mathbb{F}, k \in \mathbb{K}_{0:N_p-1},$$
$$l_i(x_i^*, u_i^*, x_i^a, x_{j \in \mathbb{I}_i}^a; 0|t) = 0,$$

which indicates that the system achieves consensus. Then according to Lemma 3, the closed-loop system achieves asymptotically consensus. This completes the proof. ∎

*Remark 6:* As discussed in Remark 3, due to the asymptotic rather than finite time consensus of assumed terminal states, the closed-loop consensus analysis in Theorem 3 becomes complex. Different from [1][8] which only use the optimal cost function in the Lyapunov function design, Theorem 3 introduces an additional infinite series term $q_i^*(t)$, which builds a bridge between finite and infinite horizon analysis to facilitate consensus analysis.

*Remark 7:* The weight matrix condition in (19) requires that an agent's self-deviation weight is greater than or equal to the summation of its out-neighbors' self-deviation weights. This condition, originally proposed in [4] and further extended in [1][8], can be explicitly guaranteed in DMPC design and can thus simplify the implementation in real applications.

## V. Applications to Multi-vehicle Cooperation

This section presents the applications of the proposed DMPC approach to multi-vehicle cooperation. Two applications including cooperative diving of multiple AUVs and platoon control of multiple CAVs are considered.

### A. Application 1: cooperative diving of multiple AUVs

In this case, we consider a swarm of AUVs following a static leading AUV for cooperative diving control. Suppose that the AUVs move in a vertical plane without yaw motion. Under the assumption of an identical constant surge velocity $u \in \mathbb{R}^+$ and a small pitch angle and heave velocity, the diving dynamics of the AUVs are simplified as [14]

$$\dot{x}_i = A_c x_i + B_c \delta_{si}, \qquad \delta_{si} \in \mathbb{D}_i,$$

$$A_c = \begin{bmatrix} 0 & -u & 0 \\ 0 & 0 & 1 \\ 0 & -\frac{z_g W - z_b B}{I_y - M_{\dot{q}}} & \frac{M_{uq} u}{I_y - M_{\dot{q}}} \end{bmatrix}, B_c = \begin{bmatrix} 0 \\ 0 \\ \frac{M_{uu\delta_s} u^2}{I_y - M_{\dot{q}}} \end{bmatrix},$$

where $x_i = [z_i, \theta_i, q_i]^T$ is the state, $z_i$, $\theta_i$, and $q_i$ are the depth, pitch angle, and pitch angular velocity of AUV $i$, respectively, $M_q$, $M_{\dot{q}}$, and $M_\delta$ are the dynamic derivative coefficients, $z_G$ and $z_B$ are locations of the center of gravity and the center of buoyancy along the z (vertical) axis with respect to the propulsion axis, respectively, $W$ and $B$ are gravity and buoyancy, $I_y$ is the moment of inertia along the y (lateral) axis, $\delta_{si}$ is the deflection angle of the stern surface, and $\mathbb{D}_i = \{\delta_i | \delta_{i,m} \leq \delta_i \leq \delta_{i,M}\}$ is the feasible control input set. The cooperative diving control objective is to coordinate the diving motion of AUVs so that they dive to a same depth and maintain a same pitch angle, i.e.,

$$\begin{cases} \lim_{t \to +\infty} e_{z,i}(t) \coloneqq z_i(t) - z_0(t) = 0, \\ \lim_{t \to +\infty} e_{\theta,i}(t) \coloneqq \theta_i(t) - \theta_0(t) = 0, \\ \lim_{t \to +\infty} e_{q,i}(t) \coloneqq q_i(t) - q_0(t) = 0. \end{cases}$$

Define a sampling time $\Delta t \in \mathbb{R}^+$, then the dynamics model is discretized as

$$x_i(t+1) = A x_i(t) + B \delta_{si}(t),$$

where $x_i(t) = [z_i(t), \theta_i(t), q_i(t)]^T$, $A = e^{A_c \Delta t}$, $B = \int_0^{\Delta t} e^{A_c t} dt \cdot B_c$, and the control objective becomes

$$\lim_{t \to +\infty} x_i(t) = x_0(t),$$

which is the same as (3).

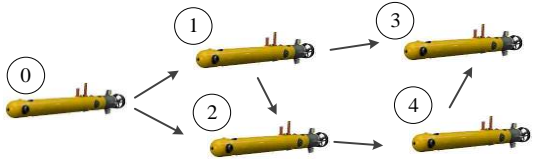

Fig. 3 Communication topology of the AUVs.

In the simulation, we consider 4 following AUVs, which have initial errors with respect to the static leading AUV that has a zero control input. The communication topology among the AUVs is shown in Fig. 3, and the other simulation parameters are given in TABLE II.

TABLE II
SIMULATION PARAMETERS OF AUVS

| Symbol | Value | Unit |
|---|---|---|
| $N$ | 4 | - |
| $x_i(0)$ | $x_0(0)=[-5.00, 0.18, 0.00]$ $x_1(0)=[-5.35, -0.42, -0.03]$ $x_2(0)=[-4.71, 0.28, -0.02]$ $x_3(0)=[-4.87, 0.43, 0.02]$ $x_4(0)=[-5.15, 0.30, 0.01]$ | [m, rad, rad/s] |
| $u$ | 0.5 | m/s |
| $M_{\dot{q}}, M_{uq}, M_{uu\delta_s}$ | -18.020, -34.192, -16.874, 10.900 | [kg m²/rad, kg m/rad, kg/rad, kg/m²] |
| $z_g, z_b$ | 0.0176, 0.0032 | m |
| $W, B$ | 497.37, 499.33 | N |
| $[\delta_{i,m}, \delta_{i,M}]$ | $[-\pi/6, \pi/6]$ | rad |
| $N_p$ | 20 | - |
| $R_i$ | 1 | - |
| $F_i$ | diag{40, 20, 4}, $i$=1 diag{10, 5, 1}, $i$=2,4 diag{0, 0, 0}, $i$=3 | - |
| $G_i$ | diag{10, 5, 1} | - |
| $K$ | [1.37; -1.94; -2.89] | - |

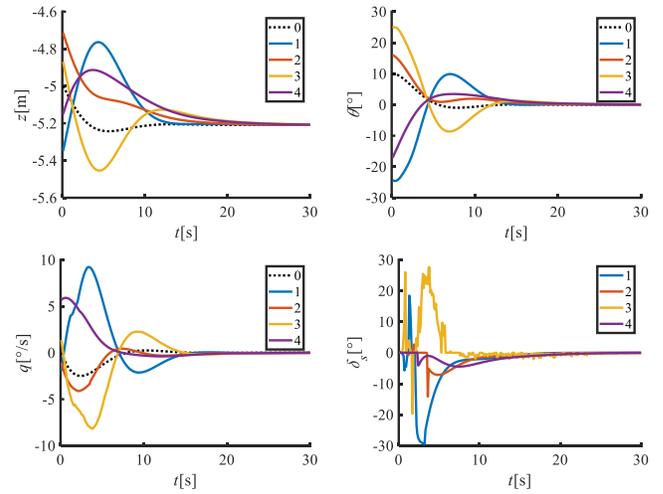

Fig. 4 State and control input profiles of the AUVs.

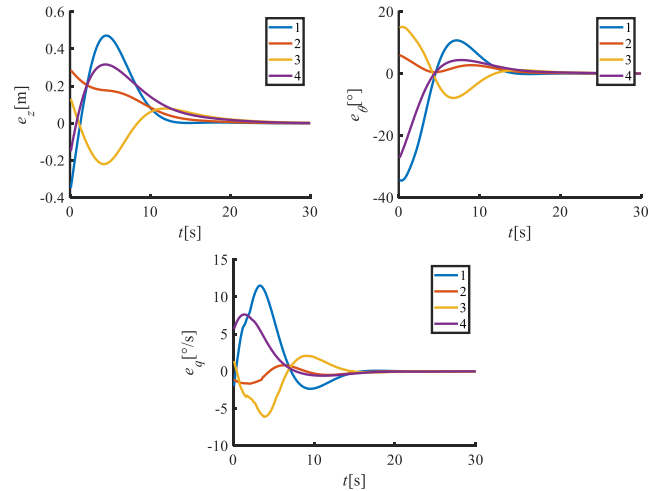

Fig. 5 Consensus error profiles of the AUVs.

As Fig. 4 and Fig. 5 show, the longitudinal motion of the following AUVs achieve asymptotic consensus with the leading AUV in the case of initial errors, and the control input constraints are also satisfied. This validates the effectiveness of the proposed theorems in the case of a static leader.

We further evaluate the impact of control input disturbance on the control performance. In this case, a random persistent disturbance signal with a magnitude of 0.1 rad and a mean value of 0 rad is imposed on each following AUV's control input. As Fig. 6 shows, the consensus errors can still converge to the neighborhood of the origin. This demonstrate the robustness of the proposed approach to control input disturbance.

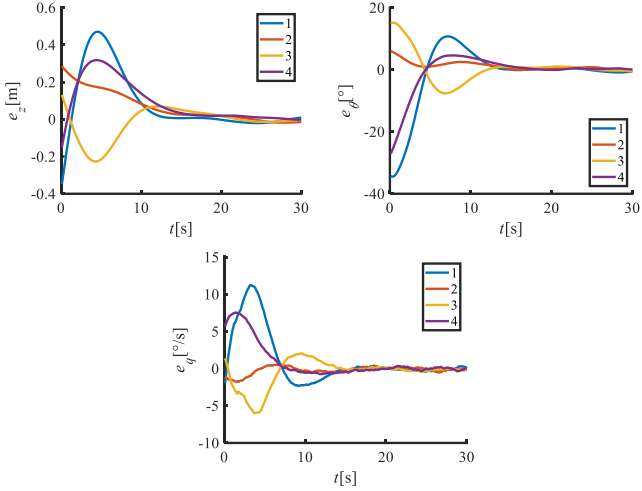

Fig. 6 Consensus error profiles of the AUVs with control input disturbance.

### B. Application 2: platoon control of multiple CAVs

In this case, we consider a platoon of CAVs following a dynamic leading CAV on a straight and flat road. The longitudinal dynamics of the CAVs are given as [2][21]

$$\dot{p}_i = v_i,$$
$$\dot{v}_i = a_i,$$
$$m_i a_i = \frac{\eta_i}{r_i} T_i - C_{A,i} v_i^2 - m_i g(f_i \cos \alpha_{r,i} + \sin \alpha_{r,i}),$$
$$\tau_i \dot{T}_i + T_i = T_{des,i},$$

where $p_i$, $v_i$, and $a_i$ are vehicle position, velocity, and acceleration, $T_i$, $\eta_i$, and $\tau_i$ are the driving torque, mechanical efficiency, and time lag of the driveline, $m_i$ and $r_i$ are vehicle mass and tire radius, $C_{A,i}$, $f_i$, and $g$ are the coefficients of aerodynamics drag, rolling resistance, and gravitational acceleration, $\alpha_{r,i}$ is the road slope. By assuming a homogeneous time lag $\tau_i = \tau$ and adopting the following feedback linearization law [21]

$$T_{des,i} = \frac{r_i}{\eta_i}\big(C_{A,i} v_i(2\tau a_i + v_i) + m_i g(f_i \cos \alpha_{r,i} + \sin \alpha_{r,i}) + m_i u_i\big),$$

where $u_i$ is a new control input representing the desired acceleration, the longitudinal dynamics can be simplified as [3][5]

$$\dot{x}_i = A_c x_i + B_c u_i, \qquad u_i \in \mathbb{U}_i,$$

$$x_i = \begin{bmatrix} p_i \\ v_i \\ a_i \end{bmatrix}, A_c = \begin{bmatrix} 0 & 1 & 0 \\ 0 & 0 & 1 \\ 0 & 0 & -\frac{1}{\tau} \end{bmatrix}, B_c = \begin{bmatrix} 0 \\ 0 \\ \frac{1}{\tau} \end{bmatrix},$$

where $\mathbb{U}_i = \{u_i | u_{i,m} \leq u_i \leq u_{i,M}\}$, $u_{i,m} \in \mathbb{R}^-$ and $u_{i,M} \in \mathbb{R}^+$ are known constants. The platoon control objective is to harmonize CAVs' motion so that they track the predecessors' velocities while maintaining a desired inter-vehicle distance $d_0 \in \mathbb{R}^+$, i.e.,

$$\begin{cases} \lim_{t \to +\infty} e_{p,i}(t) := p_i(t) - p_{i-1}(t) = d_0, \\ \lim_{t \to +\infty} e_{v,i}(t) := v_i(t) - v_{i-1}(t) = 0, \\ \lim_{t \to +\infty} e_{a,i}(t) := a_i(t) - a_{i-1}(t) = 0. \end{cases}$$

Define a sampling time $\Delta t \in \mathbb{R}^+$, then the dynamics model is discretized as

$$x_i(t+1) = Ax_i(t) + Bu_i(t),$$

where $x_i(t) = [p_i(t), v_i(t), a_i(t)]^T$, $A = e^{A_c \Delta t}$, $B = \int_0^{\Delta t} e^{A_c t}\, dt \cdot B_c$, and the control objective becomes

$$\lim_{t \to +\infty} x_i(t) = x_0(t) - [i \cdot d_0, 0, 0]^T.$$

Note that the form of the above control objective is different from that in (3). By replacing $x_j^a(k|t)$ in (9) and (11) with $x_j^a(k|t) + [(j-i) \cdot d_0, 0, 0]^T$, it is not difficult to check that the platoon control problem can be transformed into a consensus control problem and the above theorems also hold.

In the simulation, we consider 5 following CAVs, and the initial states are $x_0(0) = [0, 10, 0]^T$ and $x_i(0) = [-i \cdot d_0, 10, 0]^T$, respectively. In particular, the dynamic leading CAV moves with the following acceleration profile

$$a_0(t) = \begin{cases} 0, & t \leq 2s, \\ \sin \pi(t-5), & 2s < t \leq 6s, \\ 0, & t > 6s. \end{cases} [m/s].$$

The communication topology is given in Fig. 7, and the simulation parameters are given in TABLE III.

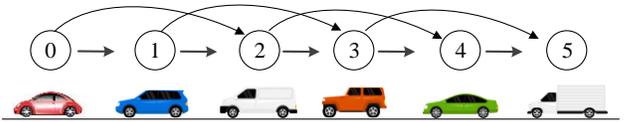

Fig. 7 Communication topology of the CAVs.

TABLE III
SIMULATION PARAMETERS OF CAVS

| Symbol | Value | Unit |
| --- | --- | --- |
| $N$ | 5 | - |
| $d_0$ | 20 | m |
| $\tau$ | 0.50 | - |
| $[u_{i,m}, u_{i,M}]$ | [-3, 3] | m/s$^2$ |
| $N_p$ | 10 | - |
| $R_i$ | 0.1 | - |
| $F_i$ | diag{20,10,4}, i=1,2,3<br>diag{5,2.5,1}, i=4<br>diag{0,0,0}, i=5 | - |
| $G_i$ | diag{5,2.5,1} | - |
| $K$ | [-0.90;-2.08;-0.96] | - |

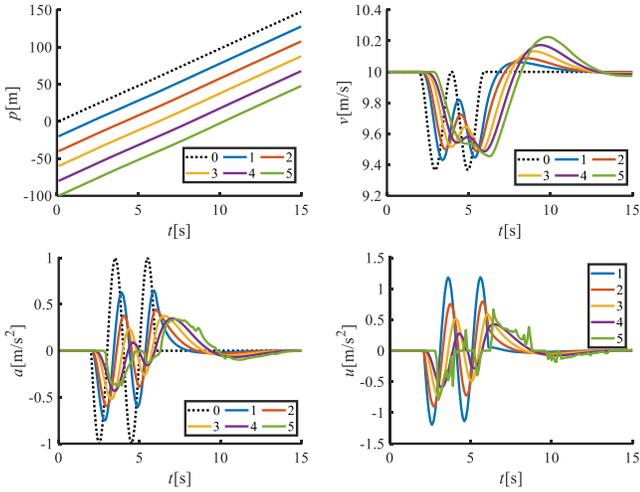

Fig. 8 State and control input profiles of the CAVs.

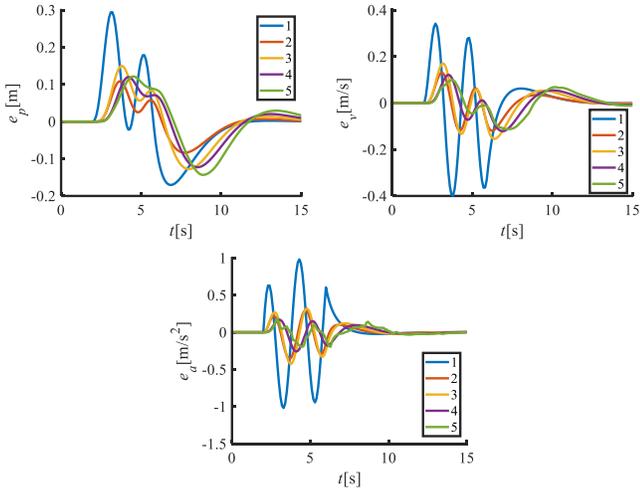

Fig. 9 Tracking error profiles of the CAVs.

As Fig. 8 shows, the following CAVs track the dynamic leading CAV well during the acceleration process. Moreover, the control input constraints are also satisfied for the following CAVs. As Fig. 9 shows, the position, velocity, and acceleration tracking errors converge to zero asymptotically as the leading CAV finishes its acceleration. This validates the effectiveness of the proposed theorems for the case of a dynamic leader. In particular, it is observed in Fig. 9 that the vehicle platoon does not seem to be strict string stable since tracking errors are not strictly attenuated during propagation along the platoon. Integrating string stability guarantee into the proposed DMPC deserves further study.

Next, we evaluate the impact of dynamics heterogeneity on the control performance. In this case, the homogeneous time lags of the 5 following CAVs are replaced with heterogeneous values [0.50, 0.38, 0.57, 0.66, 0.45], respectively, which are used by the following CAVs in predictive control. As Fig. 10 shows, the position, velocity, and acceleration tracking errors of the following CAVs can also converge to the neighborhood of the origin as the leading CAV finishes its acceleration. This demonstrates the potential of the proposed approach for heterogeneous agent dynamics.

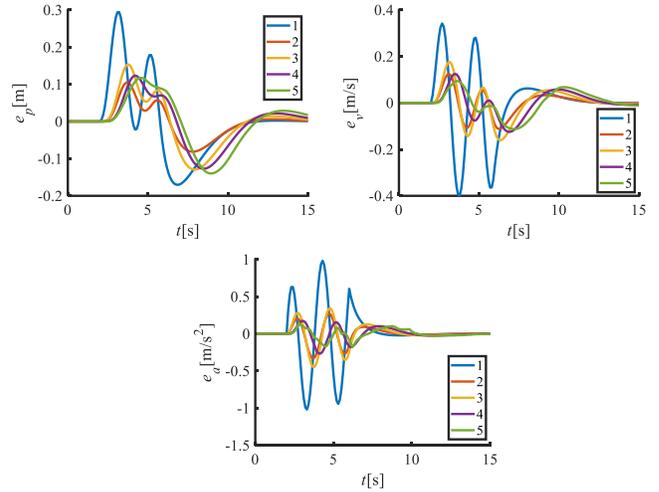

Fig. 10 Tracking error profiles of the CAVs with dynamics heterogeneity.

Finally, we evaluate the impact of both dynamics heterogeneity and model mismatch. In this case, the homogeneous time lags of the 5 following CAVs are replaced with heterogeneous values [0.50, 0.38, 0.57, 0.66, 0.45], respectively, while the nominal time lag 0.50 are used by the following CAVs in predictive control. As Fig. 11 shows, the position, velocity, and acceleration tracking errors of the following CAVs can also converge to zero as the leading CAV finishes its acceleration. In particular, compared with Fig. 9 and Fig. 10, the convergence process is slowed down. This demonstrates the robustness of the proposed approach to model mismatch.

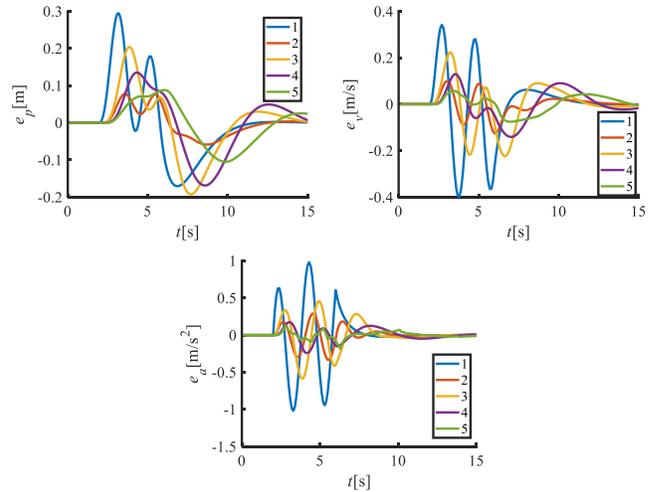

Fig. 11 Tracking error profiles of the CAVs with dynamics heterogeneity and model mismatch.

## VI. CONCLUSION

This study has proposed a DMPC approach for consensus control of multi-agent systems with general linear dynamics and bounded control input constraints. A discrete-time consensus protocol was introduced into the update lay design for assumed terminal states to guarantee terminal consensus and recursive feasibility. An infinite series of cost-to-go functions was introduced into the Lyapunov function design to guarantee closed-loop asymptotic consensus. The proposed DMPC approach was applied to two applications, i.e., cooperative control of AUVs and CAVs, which validate the effectiveness of the proposed controller.

The future work will study the effects of model uncertainties and external disturbances on the proposed approach. The impact of communication time delays and packet drops also deserve further consideration.

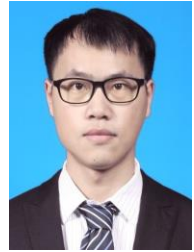

**Yougang Bian** (M'20) received the B.E. and Ph.D. degrees from Tsinghua University, Beijing, China, in 2014 and 2019, respectively. He was a visiting scholar with the Department of Electrical and Computer Engineering, University of California, Riverside, USA, from 2017 to 2018. His research interests include distributed control, cooperative control, and their applications to connected and automated vehicles. He is a recipient of the Best Paper Award at the *2017 IEEE Intelligent Vehicles Symposium*.

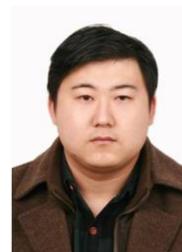

**Changkun Du** (M'20) received the B.E. degree in Automation from Henan Polytechnic University in 2008 and the M.S. and Ph.D. degrees in Control Science and Engineering both from Beijing Institute of Technology in 2014 and 2019, respectively. From 2017 to 2018, he was a visiting scholar in the Department of Electrical and Computer Engineering, University of California, Riverside, USA. He is currently a postdoctor at Beijing Institute of Technology, China. His research interests include multi-agent systems, event-triggered control, finite-time control, and control of connected vehicles.

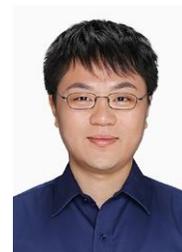

**Manjiang Hu** received the B.Tech. and Ph.D. degrees from Jiangsu University, China, in 2009 and 2014, respectively. He worked as a postdoc in the Department of Automotive Engineering, Tsinghua University from 2014 to 2017. He is currently a professor with the College of Mechanical and Vehicle Engineering, Hunan University, Changsha, China. His research interests include cooperative driving assistance technology and vehicle control.

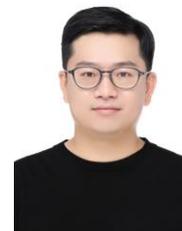

**Haikuo Liu** (M'20) received the B.Sc. degree from the Henan Normal University, Henan, China, and the Ph.D. degree from Beijing Institute of Technology, Beijing, China, in 2013 and 2020, respectively. He is currently a postdoc at School of Mechatronical Engineering, Beijing Institute of Technology, Beijing, China. His research interests include cooperative control of multi-agent systems, event-triggered control, and their applications on connected vehicles and vibration control of large structures.